\documentclass{revtex4}
\usepackage{graphicx} 
\usepackage{amsmath,amssymb}
\usepackage{bm,bbm}

\usepackage[colorlinks,linkcolor={blue},citecolor={blue}, urlcolor={blue}]{hyperref}

\usepackage{xcolor}

\begin{document}
\title{Chirality Effects in Molecular Chainmail}
\author{Alexander R. Klotz}
\affiliation{Department of Physics and Astronomy, California State University}
\author{Caleb J. Anderson}
\affiliation{Department of Physics and Astronomy, California State University}
\author{Michael S. Dimitriyev}
\affiliation{Department of Materials Science and Engineering, Texas A\&M University}

\begin{abstract}
Motivated by the observation of positive Gaussian curvature in kinetoplast DNA networks, we consider the effect of linking chirality in square lattice molecular chainmail networks using Langevin dynamics simulations and constrained gradient optimization. Linking chirality here refers to ordering of over-under versus under-over linkages between a loop and its neighbors. We consider fully alternating linking, maximally non-alternating, and partially non-alternating linking chiralities. We find that in simulations of polymer chainmail networks, the linking chirality dictates the sign of the Gaussian curvature of the final state of the chainmail membranes. Alternating networks have positive Gaussian curvature, similar to what is observed in kinetoplast DNA networks. Maximally non-alternating networks form isotropic membranes with negative Gaussian curvature. Partially non-alternating networks form flat diamond-shaped sheets which undergo a thermal folding transition when sufficiently large, similar to the crumpling transition in tethered membranes. We further investigate this topology-curvature relationship on geometric grounds by considering the tightest possible configurations and the constraints that must be satisfied to achieve them.
\end{abstract}

\maketitle

\section{Introduction}
Many emerging materials are comprised at the microscopic scale of two-dimensional crystalline materials~\cite{geim} or topologically complex linked-ring molecular architectures such as polycatenanes~\cite{wu2017poly} and Olympic gels~\cite{liu2023structurally}. Some materials have combined both features into molecular structures that are both planar and topologically linked~\cite{august2020self, thorp2015infinite}. There is currently an incomplete understanding about how the underlying topology of the molecular network affects its equilibrium mechanical properties, as well as what role thermal fluctuations play in the stability of planar molecules at finite temperatures. One emerging model system for studying these effects is kinetoplast DNA.

Kinetoplasts are the mitochondrial DNA of trypanosome parasites. Often described as molecular chainmail, they consist of several thousand topologically linked DNA ``minicircles'' of a few thousand base pairs forming a planar network as well as several dozen ``maxicircles'' of tens of thousands of base pairs (similar to our own mitochondrial DNA) also linked within the network~\cite{shapiro1995structure}. Their catenated (linked-ring) molecular structure and planar network topology make them an interesting experimental system for researchers in both the topological chemistry and two-dimensional materials communities.

In one of the initial studies characterizing the material properties of kinetoplasts in free solution, Klotz et al.~reported positive Gaussian curvature, giving kinetoplasts the appearance of wrinkled hemispheres~\cite{klotz2020equilibrium}. There is no known biological reason for this curvature, and it is likely not present in every species. Subsequent work attempted to explain this curvature on physical grounds. Simulations by Polson, Garcia, and Klotz~\cite{polson2021flatness} showed that networks of rigid linked rings, when thermalized, will replicate the curvature seen in \textit{Crithidia fasciculata} kinetoplasts. In a detailed simulated-supported atomic force microscopy investigation of kinetoplast DNA, He et al.~\cite{he2023single} argued that the incommensurability between the area of the kinetoplast sheet and its circumference, which is lined by a dense fiber of excess DNA linkages, creates a tension that is resolved by curving the membrane.

Based on gel electrophoresis experiments~\cite{chen1995topology}, it was concluded that the network structure of \textit{Crithidia} kinetoplasts is that of a honeycomb lattice, with the average number of minicircles each minicircle is linked to, termed the valence, most likely being three. Subsequent investigations, including that of He et al.~\cite{he2023single}, supported the average trivalence but not the honeycomb picture, which cannot capture the observed complexity of minicircle arrangement. More recent work has investigated the role of the edge loop in determining kinetoplasts topology~\cite{ragotskie2023effect}, as well as that of the maxicircles that are typically neglected~\cite{ramakrishnan2023single}. 
The dependence of the equilibrium structure of a molecular chainmail network on its lattice topology has not been fully explored.
Preliminary work by Thomas O'Connor showed that different medieval chainmail designs folded differently at low temperatures~\cite{thomasconference}, while the simulations by Polson, Garcia and Klotz showed slight differences in the concavity of hexagonal and square networks with non-uniform valence~\cite{polson2021flatness}. 
The most detailed simulations by He et al.~\cite{he2023single} used trivalent honeycomb networks, and highlighted the role of link \textit{chirality} as well as network topology.

\begin{figure}
    \centering
    \includegraphics[width=0.5\textwidth]{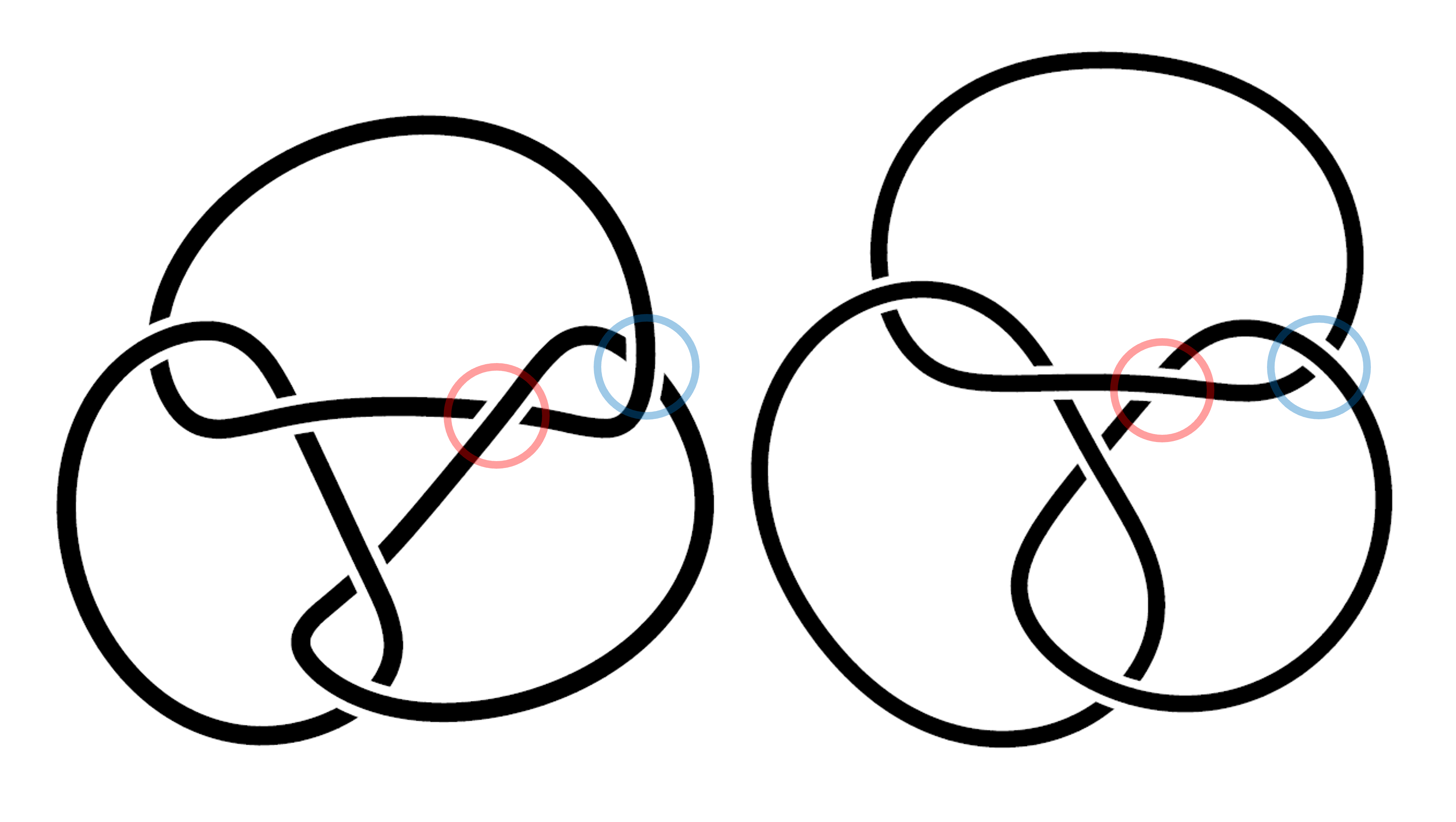}
    \caption{Two three-loop six-crossing links that differ by the chiralities of the linkages in the red and blue circles. The link on the left ($6_{1}^{3}$) is alternating, while the link on the right ($6_{3}^{3}$) is non-alternating. They can be interconverted by cutting the loop between the circles, sliding the two ends over and under the loop they were linked to, and refastening them.}
    \label{fig:sixlinks}
\end{figure}

When two loops share a topological Hopf link, a given projection may show one link passing over then under the other, or under then over. But, either loop may be rotated about the plane of projection, changing an over-under to an under-over. When a loop is topologically linked with several others, each with their own topological constraints, there are non-equivalent choices of how the loop in question passes over or under its neighbors. Consider for example the two three-ring, six-crossing links depicted in Fig.~1. The link on the left is alternating: starting at some point on each loop and travelling clockwise or counterclockwise, the linking with the neighbors can be read as ``over-under-over-under'' or ``under-over-under-over.'' The link on the left can be transformed into the link on the right by cutting one of the loops, twisting the cut loop such that the remaining connection changes sign, and reconnecting the loop around its former neighbor. The link on the right is non-alternating, meaning certain loops will have adjacent over-over or under-under crossings with their neighbors. A simple three-link chain does not admit these two chiralities, it is the closure into a triangle that introduces this constraint, and this applies to every loop in a chainmail network. We note that the term ``chirality'' here does not have the same meaning as is used in knot theory and stereochemistry. The over-under and under-over projections represent two chiralities of the oriented Hopf link, and we refer to the different link topologies that can be constructed by varying the link chirality of each connection. The two links in Fig.~1, however, are not mirror images of each other.

The effect of link chirality in polycatenanes and molecular chainmail was explored by Luca Tubiana and collaborators in two studies. The first examined twisted circular polycatenanes~\cite{tubiana2022circular}, finding that a closed, untwisted, fully non-alternating chain-link will be structurally relaxed at equilibrium, but adding twists before closing the chain link will introduce a torsional constraint to the system that must be alleviated. They formulated a relationship between the number of twists and the apparent writhe of the polycatenane that was analogous to the Fuller-White-Calugareanu relation that applies to supercoiled DNA~\cite{dennis2005geometry}. In a second study, the simulations used by He et al.~to explore the effect of tension due to the kinetoplast edge loop~\cite{he2023single}, randomized the chirality of each loop such that effects of chirality were averaged out. During the writing of this manuscript, we became aware of parallel work by Luengo-M\'arquez et al.~\cite{luengo} examining the equilibrium properties of block-copolymer chainmail networks with two different link chiralities. The networks simulated by Polson, Garcia, and Klotz~\cite{polson2021flatness} used ``Japanese-style'' networks in which tetravalent or hexavelent loops on a lattice were connected by divalent linkers, eliminating issues with link chirality.

Here, we consider the effect of molecular chirality in square lattices of topologically linked loops. 
Although the honeycomb lattice is typically used to describe kinetoplast DNA, we choose a square lattice, as the bipartite structure ensures a consistent choice of linking chirality, i.e.~without the ``geometric frustration'' characteristic of non-bipartite lattices.
Additionally, the 4-valent connectivity yields a bigger difference between the fully alternating and fully non-alternating cases. The three square lattice chiralities that we study can be seen in Fig.~2. Examining a given loop and its linkages with its neighbors, we may start in the first panel at ``12 o'clock'' and read the crossings clockwise as ``under-over-under-over-under-over.'' This is the alternating configuration. The second panel may be read as ``under-under-over-over-under-under-over-over,'' which we term the fully or maximally non-alternating configuration. The third panel may be read as ``under-under-over-under-over-over-under-over,'' which we refer to as semi-alternating. The semi-alternating pattern is known in the armoring community as the European 4-in-1 weave~\footnote{Historical armor styles mentioned in this paper are based on the conventions of online communities. We have not examined the historical accuracy of their names.}. The unit cells of these networks are not the individual loops but rather the junctions of four loops. In Alexander-Briggs and Thistlethwait notation, the unit cells of each network are  $8_{1}^{4}$/L8a21, $8_{3}^{4}$/L8n8, and $8_{2}^{4}$/L8n7.   Underneath each example is a rendering of a $3 \times 3$ network with that chirality. Each pair of crossings may be treated as an edge on a square lattice. If we assign clockwise over-under crossings with a black dot and clockwise under-over configurations with a white dot, the square lattices within each cell show the tilings associated with each topology. These tilings bear similarity to the system of elastic defects studied by Plummer et al.~\cite{plummer2020buckling}, in which defects could rise up or down out of the plane of a lattice of springs, and equilibrated to ``ferromagnetic,'' and the ``antiferromagnetic'' striped and checkerboard configurations similar to our three chiralities.  As each pattern of elastic defect lattice introduced different features in their simulated surfaces, we will see similar features arise in molecular chainmail.

\begin{figure}
    \centering
    \includegraphics[width=\textwidth]{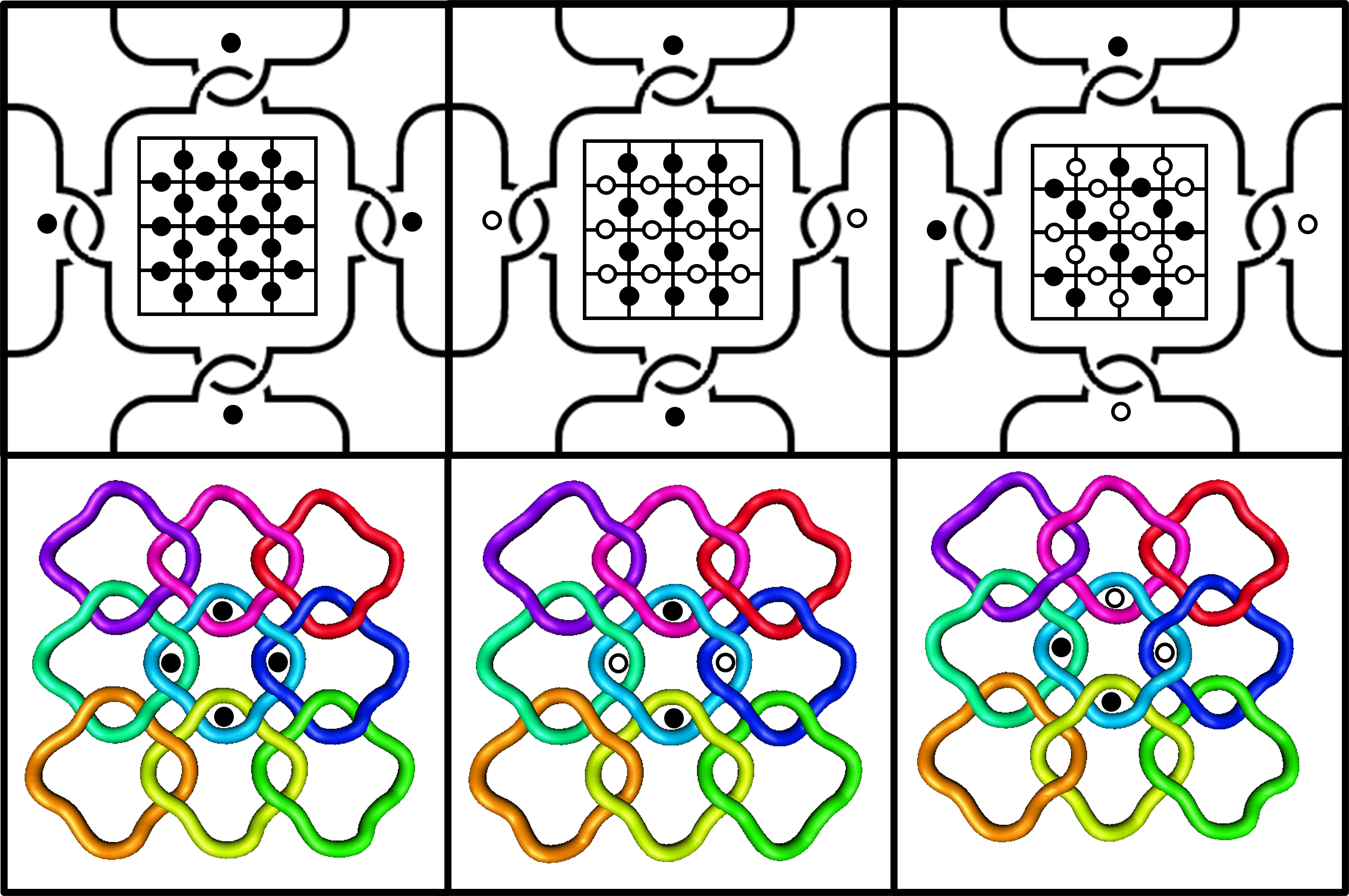}
    \caption{The linking of a single loop with its four neighbors in the alternating (left), fully non-alternating (middle), and semi-alternating (right) chiralities. Within each loop is a $4 \times 4$ network tiled by these chiralities, with black dots representing over-under crossings and white dots representing under-over. The bottom shows renderings of the initial conditions of $3 \times 3$ lattices of each chirality.}
    \label{fig:enter-label}
\end{figure}

In this work, we explore the configurations of molecular chainmail given these chiral topological constraints. We use two complementary numerical methods and a simplified exactly-solvable model. The first numerical method is Langevin Dynamics (LD), used to model the equilibrium behavior of polymer chains. We explore the equilibrium configurations of chainmail networks parameterized to the properties of DNA in low ionic strength solvents, as explored in fluorescence experiments with kinetoplast DNA~\cite{soh2020ionic}. The equilibrium configuration of a polymer depends on a balance of entropy, excluded volume, and bending rigidity. To reduce the degrees of freedom more directly investigate the relationship between topology and geometry, the second method uses constrained gradient optimization (CGO) to find the tightest possible configuration of chainmail networks, which might arise in synthetic chemical networks~\cite{thorp2015infinite}. This algorithm is typically used to find the tightest configurations of knots and links, and here is used to identify the minimal geometric factors that lead to the observed membrane behavior. Finally, we develop a simple model of the preferred homogeneous embedding of these networks, illustrating the emergence of intrinsic curvature and the necessity of ring deformation due to an apparent geometric incompatibility between certain periodic linkages and constraints imposed by Euclidean space.

\section{Methods}

\subsection{Initialization}
We initialize our loops (a term we use interchangeably with links to refer to the individual components of the networks) as 16-gons. These 16-gons are centered on square lattice sites, and can be thought of as inhabiting tiling squares. Eight vertices of each 16-gon lie in the $xy$-plane in an octagon, with four sides touching neighboring squares in the lattice. Between each two octagonal vertices on the sides of the square, are points extending into the next square, one in the $+z$ direction and one in the $-z$ direction. The order of the $\pm z$ extension determines whether the linkage is over-under or under-over. Each 16-gon has four linkage components, the alternating 16-gons have all four going over-under. The $z-$coordinates on opposite sides of the 16-gon are flipped to create the fully non-alternating network. To create the semi-alternating network, the $z-$coordinates are flipped on either the upper and right linkage, or the lower and left linkage, and the network is created by a checkerboard tiling of those two cases. Schematics of $3 \times 3$ initial configurations can be seen in Fig.~2.

We use spline interpolation to refine the mesh for each loop, increasing the number of vertices-per-loop to $\sim 24$.
Because the algorithms we use typically function better with smooth initial configurations, we import the generated configurations into KnotPlot~\cite{scharein} and anneal them by treating each vertex as a charge and each edge as a spring, which ensures evenly-spaced vertices and a safe distance between beads on neighboring links such that they will not cross each other during simulation. After the rings in a medium-sized network have been annealed in KnotPlot, we can extract the coordinates of an interior ring and tile them to create larger networks. Coordinates of these rings and instructions for tiling them are available as ancillary data. For CGO, we also reduce the charge of each vertex and apply a contour-minimizing force in KnotPlot, which allows the minimization to proceed faster.

\subsection{Langevin Dynamics}

We simulate molecular chainmail networks with $M$ links using a model used to simulate topologically complex polymers in previous works, and our descriptions may bear similarity to previous descriptions of these methods. In short, each loop in the network is comprised of beads of diameter $\sigma$ (which sets the lengthscale of the system) at position $\mathbf{r}_i(t)$, connected by springs to their two neighbors. A finitely-extensible nonlinear elastic (FENE) spring potential with a maximum extension of $1.5 \sigma$ is used. Excluded volume interactions between beads are enforced by a truncated Lennard-Jones repulsive potential that applies when the centers of mass of two beads are closer than $\sigma$. The relatively short range of distances between the excluded volume of the beads and maximum extension of the springs ensures that strands do not cross and the link topology is preserved. Bending rigidity is imposed by a Kratky-Porod potential depending on the cosine of the angle between three successive beads. The strength of this potential sets the persistence length of the polymer. The entire contribution to the energy of a bead is:
\begin{equation}
    U_{\rm tot}=U_{\rm spr}+U_{\rm ev}+U_{\rm bend}\, .
\end{equation}
The excluded volume interaction takes the form:
\begin{equation}
    U_{\rm ev}=\begin{cases}
  4\epsilon\left[\left(\frac{\sigma}{r}\right)^{12}-\left(\frac{\sigma}{r}\right)^{6}+\frac{1}{4}\right]
    & \text{if } r\leq 2^{1/6}\sigma\\
    0              & \text{otherwise},
\end{cases}
\end{equation}
where $\epsilon$ sets the energy scale of the repulsive interactions. The spring force is parameterized as:
\begin{equation}
    U_{\rm spr}=\begin{cases}
    -\frac{1}{2}\left(\kappa\frac{\epsilon}{\sigma^2}\right)R_{\rm max}\log\left| 1-\left(\frac{r}{R_{\rm max}}\right)^{2}\right|,  & \text{if } r\leq R_{\rm max}\\
    0             & \text{otherwise},
\end{cases}
\end{equation}
where $\kappa$ is typically 30 and sets the spring constant in units of $\epsilon/\sigma^2$ and $R_{\rm max}$ is the maximum separation of the springs, and is 1.5$\sigma$ in this work. The bending potential takes the form:
\begin{equation}
    U_{\rm bend}=\frac{\ell_{p}}{\sigma}kT(1-\cos{\theta})\,.
\end{equation}
The dimensionless ratio of the persistence length $\ell_{p}$ to the bead diameter is typically $\ell_p/\sigma = 5$ in this work. The time evolution of the $i$\textsuperscript{th} bead is determined by the Langevin equation:
\begin{equation}
    m\ddot{\mathbf{r}}_i(t) =-\gamma\dot{\mathbf{r}}_i(t) - \nabla_{\mathbf{r}_i} U_{\rm tot}+\sqrt{2kT\gamma}\,\bm{\eta}(t)\,.
\end{equation}
Here, $\gamma$ is the drag coefficient on a single bead, $kT$ is the thermal energy scale, $\bm{\eta}$ is a delta-correlated normal random variable, i.e.~$\langle \eta_i(t) \eta_j(t') \rangle = \delta_{ij}\delta(t-t')$, and an overdot represents a time derivative. The final term provides a random force that emulates Brownian motion in a manner consistent with the fluctuation-dissipation theorem. These equations of motion are solved by LAMMPS~\cite{lammps}, which iterates the system forward in time using the Velocity Verlet algorithm.

The system is non-dimensionalized with $\sigma$, $\gamma$, $m$, $kT$ and $\epsilon$ taking values of 1, which defines a timescale $\tau=\sigma\sqrt{m/\epsilon}$. We iterate the simulation with a timestep of 0.01 $\tau$. We initially perform 500 iterations of the system with a simpler harmonic spring potential to avoid overstretched FENE springs. We then iterate the system for hundreds of thousands to tens of millions of timesteps depending on the system size, typically at least ten times as long as the initial transient deformation of the network from its initial conditions. For each system size, we simulate at least five iterations with different random seeds. To ensure the harmonic equilibration stage did not change the linking topology through strand crossing, we compute the Gauss linking number of neighboring loops in the final configuration and verify that everything is linked appropriately. In practice, this was only an issue for unfavorable initial conditions, which were fixed in KnotPlot. 

The equilibrium configuration of a polymer chainmail network depends on four lengthscales: the effective width of the monomers, the persistence length, the contour length of each ring, and the total number of rings. In our simulations, we primarily use a parameterization in which the effective width is set by the diameter of each bead, the persistence length is five beads (representing a monomer anisotropy found in DNA in low-salt solutions), each ring is 24 beads (about 5 persistence lengths), and the number of rings is varied. The simulated polymer rings are comparable to the minicircles of \textit{Trypanosoma brucei}, which are slightly below six persistence lengths in contour. The minicircles in \textit{Crithidia fasciculata} are each about 16 persistence lengths in contour, requiring 80 monomers per ring in this model. While we primarily focus on the system with M rings of 24 beads and a persistence length of 5$\sigma$, we will discuss the effect of varying the other lengthscales.

\subsection{Constrained Gradient Optimization}

We use constrained gradient optimization (GCO) to find the tightest configuration of molecular chainmail networks, that which minimizes the total contour length of all the loops while treating each loop as a tube of unit radius and respecting a no-overlap constraint. The width of a curve is defined as the minimum radius of a circle that is guaranteed to pass through any three points on a curve. The ropelength of a knotted or linked curve is the minimum ratio of the total contour length of the curve to its width. Conventionally, this is normalized such that the thickness of the curve is 0.5 or 1, depending on whether it is treated as a rope with unit diameter or radius. The configuration of a knot or link that minimizes ropelength is known as ideal. No exact value for the ropelength of a nontrivial knot is known, but numeric analyses put the upper bound of the ropelength of the trefoil knot at around 32.74 radii \cite{ashton2011knot}.

To perform constrained gradient optimization (CGO) on chainmail networks, we use an algorithm called Ridgerunner, developed by Jason Cantarella and colleagues. A full description of Ridgerunner can be found in Ashton et al.~\cite{ashton2011knot}, but in short it approaches the ideal configuration of an initial knot by perturbing its coordinates into many trial knots. At each step, the next perturbation is computed by projecting the gradient of the length function onto a polyhedral cone of perturbations of the current polygon which respect the no-overlap condition. When the projected gradient is a small fraction of the length of the original gradient vector (typically 0.01 or 0.001) the algorithm terminates.

We initialized the three types of square lattice chainmail networks with 9, 16, and 25 loops, with 16 vertices per loop. As a preliminary investigation we also tightened twisted polycatenanes, the system established by Tubiana et al.~\cite{tubiana2022circular} to study chirality effects in linked polymers, discussed in the appendix. We initially performed CGO on the networks using an equilateralization force to maintain a constant distance between the vertices of each loop. When the residual gradient reached 0.1 or the ropelength stopped decreasing, the final configuration was re-run without equilateralization, towards its minimum. Compared to Langevin Dynamics and similar methods, CGO is generally slower and limited in system-size as width-testing scales cubically with the number of nodes, and is susceptible to strong local minima. Ridgerunner is not optimized for systems with configurations that have both straight and tightly curved components, such as the links in chainmail networks. 
Thus, our results should be regarded as approximations to length-minimizing links, within the constraints discussed above.

\section{Results and Discussion}

\subsection{Langevin Dynamics}

Our primary investigation simulated square lattices with the alternating, fully non-alternating, and semi-alternating chiralities, from $3\times3=9$ loops up to $15\times15=225$. We also simulated a more circular network with 137 loops, and performed trial simulations of square Japanese-style chainmail and Borromean chainmail (Appendix). Alternating and fully non-alternating networks were simulated with $M = 9$, 16, 25, 49, 81, 137, and 225. The behavior of the semi-alternating networks proved more complex, and thus they were more densely sampled at $M = 9$, 16, 25, 36, 49, 64, 81, 121, 137, 169, 196, and 225.

\begin{figure}
    \centering
    \includegraphics[width=\textwidth]{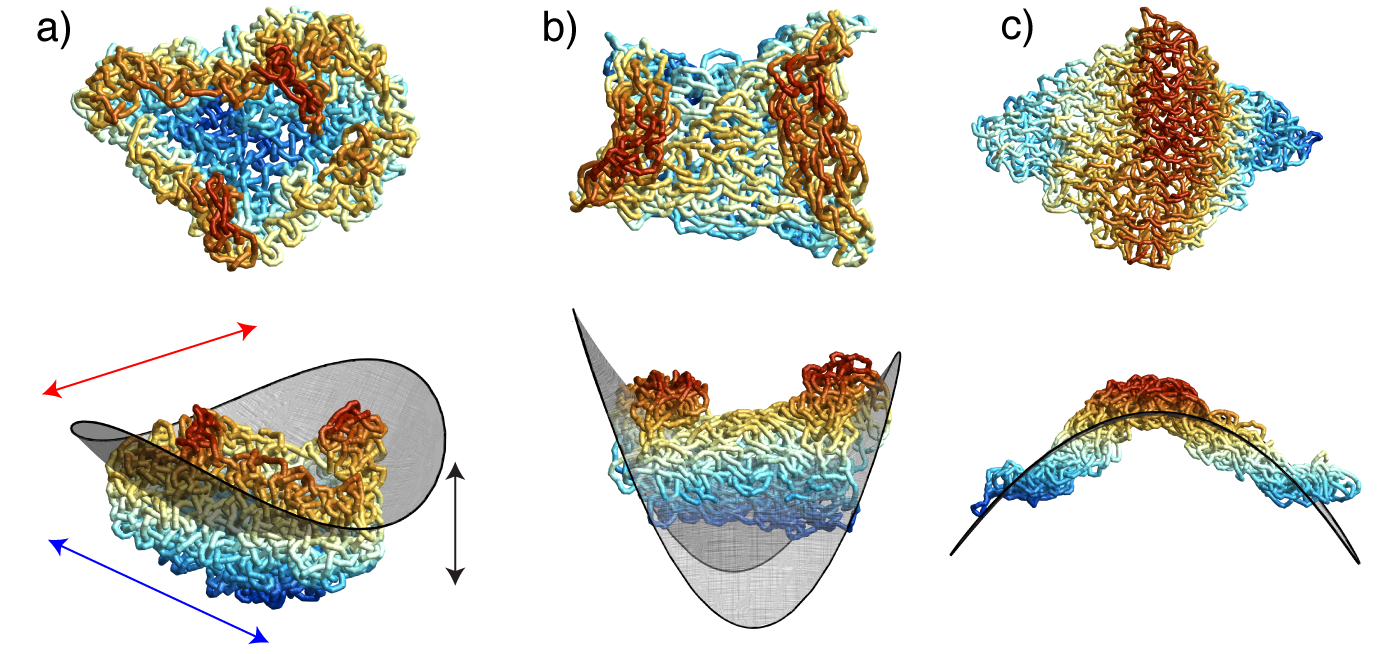}
    \caption{Representive images of 225-loop molecular chainmail networks simulated with Langevin Dynamics for a) alternating, b) fully non-alternating, and c) semi-alternating chiralities. Top row shows a top-down orientation, bottom row shows a side view with an osculating surface. Links are color-coded by their distance along the surface normal direction. The arrows in the bottom-left image represent the principal axes of the gyration tensor.}
    \label{fig:postlammps}
\end{figure}

We begin with qualitative descriptions of the simulated networks, examples of which can be seen in Fig.~\ref{fig:postlammps}. Videos of three-dimensional rotations of each case may be found in the ancillary data. We observe that alternating networks become bowl-shaped with positive Gaussian curvature, similar to \textit{Crithidia} kinetoplasts. Fully-alternating networks become saddle-shaped with negative Gaussian curvature. Semi-alternating networks appear flat, although can fold along one axis when sufficiently large, discussed subsequently. In the first two cases, the square geometry of the lattice leaves no strong imprint on the final configuration, and the edges of alternating configurations resemble the typical edge shapes seen in kinetoplast DNA~\cite{klotz2020equilibrium}. The semi-alternating lattices quickly elongate along one pair of opposite corners and contracting along another, converging on a diamond configuration with an aspect ratio of about two.

\begin{figure}
    \centering
    \includegraphics[width=0.7\textwidth]{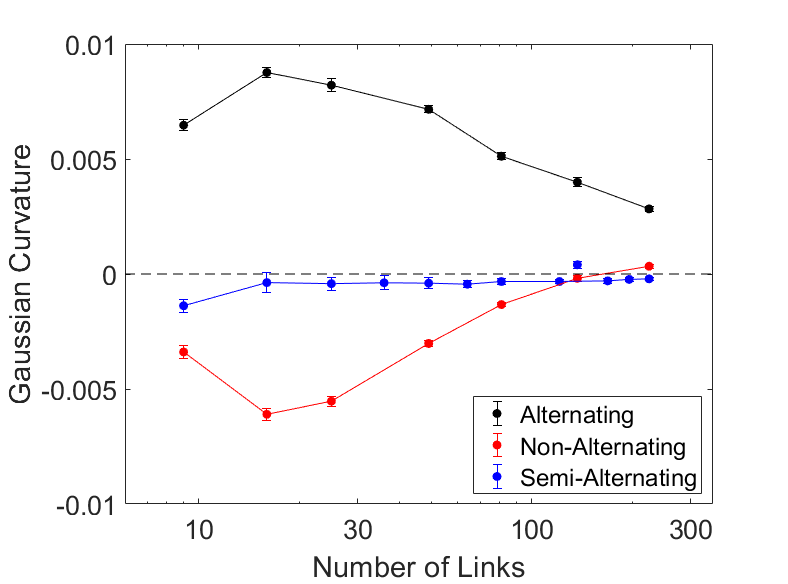}
    \caption{Gaussian curvature as a function of network size for alternating (black), fully non-alternating (red) and semi-alternating (blue) networks. The datum at $M=137$ in each set was taken from a circular rather than square network.}
    \label{fig:gauss}
\end{figure}

In order to obtain a measure of curvature of each chainmail assembly, we first need to extract a suitable representation of a surface from a given assembly.
Here, we generate a Delaunay triangulation of center of mass of each ring, based on the square lattice connectivity.
This creates a membrane of plaquettes that meet at nodes, each of which has a local Gaussian and mean curvature that can be computed from the algorithm laid out by Meyer et al.~\cite{meyer}. The Gaussian curvature is averaged over all nodes to characterize the curvature of each sheet. This value has units of 1/$\sigma^2$, and has a reciprocal on the order of the squared radius of gyration for curved networks. The surface- and time-averaged Gaussian curvature of the networks as a function of their size can be seen in Fig.~\ref{fig:gauss}. The positive curvature of the alternating networks and negative curvature of the non-alternating networks mirror each other, both reaching a peak at $M=16$ before approaching zero. 
The maximum at $M = 16$ can be regarded as a balance between two competing trends: for small $M$, the polymer rings are subject to fewer constraints and are thus able to undergo larger conformational fluctuations, reducing the orientational correlations needed to establish a well-defined curvature; for larger $M$, bending modes of the membranes are more easily excited by thermal fluctuations, leading to inhomogeneous curvature distributions that reduce the average curvature.
The semi-alternating networks have a curvature much closer to zero than the others, as expected, but is consistently negative at about $-0.0003\sigma^{-2}$ rather than fluctuating about zero, approaching zero with system size. The notable exception is the circular sheet at $M=137$. While the averaged Gaussian curvature of the largest non-alternating network passes zero, visual inspection reveals it to still be saddle shaped (Fig.~3b).
This is confirmed by an alternative surface-construction procedure in which the networks are fit to quadric surfaces, as illustrated in the lower panels of Fig.~\ref{fig:postlammps} and discussed in the Appendix.

The parallel work by Luengo-M\'arquez et al.~\cite{luengo} examined two chiralities of the honeycomb lattice and also observed negative Gaussian curvature in systems equivalent to our non-alternating chirality. They examined the local variation of Gaussian curvature across the membrane, and their analysis showed that the time-averaged curvature at each vertex on the surface matched the local curvature of the time-averaged configuration, suggesting that the curvature is stable over long time periods.

To characterize the dimensions of the equilibrated networks, we compute the gyration tensor based on the position of each of $N$ beads relative to the center of mass:
\begin{equation}\label{eq:gyration}
    G_{ab}=\frac{1}{N}\sum_{i=1}^{N}(r_{i,a}-\langle r_{a}\rangle)(r_{i,b}-\langle r_{b}\rangle),
\end{equation}
where $r_{i,a}$ and $r_{i,b}$ represent the position of the $i$\textsuperscript{th} bead in dimensions $a$ and $b$, and the angle brackets denote the center of mass in that dimension. The eigenvalues of this tensor represent (squared) lengthscales describing the distribution of the network along three perpendicular axes, the eigenvectors corresponding to each eigenvalue. The square roots of these eigenvalues are the principal radii of gyration, and when ordered by size are referred to as the minor, medium, and major axes. For sheet-like networks, there are typically two eigenvectors that point within the plane of the sheet with comparable eigenvalues, and a third pointing transverse to the sheet with an eigenvector that scales with a weaker exponent than those in the plane, sometimes called the roughness exponent. This is known as the ``flat'' phase~\cite{kantor1993excluded}, as membranes will extend asymptotically farther in their chemically-defined plane than fluctuate transverse to it. In the flat phase, the in-plane gyration eigenvalues typically grow close to linearly with molecular weight (consistent with area-mass proportionality). Scaling arguments put the roughness exponent near 0.7~\cite{kroll1993floppy}, which is consistent with simulations~\cite{popova2007structure, mizuochi2014dynamical, gandikota2023crumpling}. The simulation of chainmail networks of rigid circles by Polson et al.~\cite{polson2021flatness} found in-plane exponents between 0.92 and 1.02 and roughness exponents between 0.73 and 0.84, likely higher than the tethered membrane flat phase due to curvature effects. The power-law scaling parameters for each eigenvalue can be found in Table~1. Here we report the dependence on the squared principal radius scaling with molecular weight; other literature may report either the radii or their squares, and either the molecular weight or the side-length, leading to two possible factor-of-two differences in how the exponents are presented (e.g.~the major eigenvalue in the flat phase is reported in various studies to be close to 0.5, 1, and 2).

\begin{table}[]
\begin{tabular}{|l|l|l|l|}
\hline
Eigenvalue & Alternating & Non-Alternating & Semi-Alternating \\ \hline
Minor & $0.66\pm0.02$ & $0.74\pm0.02$ &  $0.44\pm0.02$ \\ \hline
Medium & $0.79\pm0.02$ & $0.73\pm0.01$ &  $0.93\pm0.01$ \\ \hline
Major & $0.76\pm0.02$ & $0.71\pm0.03$ & $0.95\pm0.01$ \\ \hline
\end{tabular}
\caption{Best-fit scaling exponents of gyration tensor eigenvalues.}
\end{table}

\begin{figure}
    \centering
    \includegraphics[width=\textwidth]{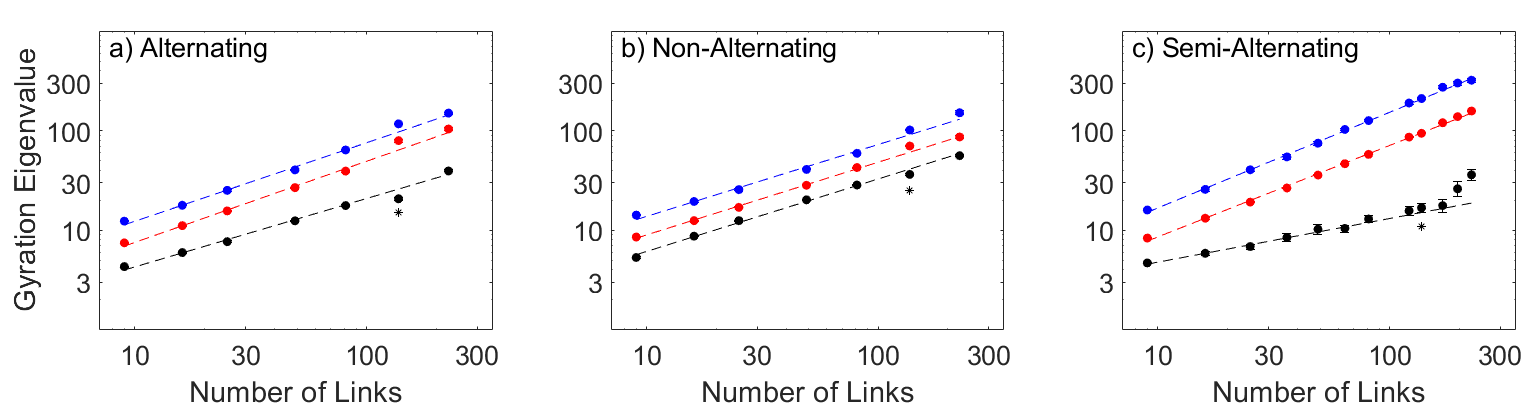}
    \caption{Squared principal radii of gyration from the eigenvalues of the gyration tensor for alternating (a), full non-alternating (b) and semi-alternating (c) networks as a function of molecular weight, with power-law fits. Each eigenvalue is labelled black, red, and blue in increasing size. The asterisk indicates the 137-ring circular networks that were not used for fits. If not visible, error bars are smaller than points.}
    \label{fig:rgs}
\end{figure}

The characteristic squared radii of each chirality behave differently with the number of links, and we will discuss them individually. The positively-curved alternating networks have middle and major squared radii that scale with a size exponent between 0.76 and 0.79. This is lower than what is typically found in tethered membrane simulations, including the linked-ring simulation study~\cite{polson2021flatness}. The minor exponent of about 0.66 is comparable to some simulations of the flat membrane phase, but should be interpreted as representative of curvature, e.g.~the depth of the bowl shape adopted by the membrane, rather than roughness. The minor exponent is about $5/6$ that of the major exponents, which is comparable to what was observed in the linked ring simulation for rings of finite thickness. The fact that the depth of the bowl grows with a weaker exponent indicates that asymptotically large networks may not appear significantly curved. If, ignoring wrinkles and fluctuations, the membrane took the form of a spherical cap with constant solid angle, all eigenvalues would scale similarly. The smaller transverse scaling is consistent with this angle decreasing with length, or equivalently, the effective radius of the osculating sphere increasing faster than the dimensions of network size. We note that although there is no reason to suspect that kinetoplast DNA has alternating chirality, simulating alternating molecular networks may be a useful method of inducing curvature in future kinetoplast simulation studies. 

The negatively-curved, fully-alternating networks have three principal squared radii that scale with roughly the same exponent, between 0.70 and 0.75. This indicates that they asymptotically adopt an isotropic configuration. 
Negatively curved saddle-shaped membranes are not common in nature or in synthetic materials, but it is possible to induce negative curvature in origami sheets with specific folding patterns~\cite{sardas2024continuum}. 
There have been predictions of nontrivial electronic effects in negatively curved graphene~\cite{kolesnikov2008electronic}, but this has not been synthesized. In future catalysis systems it may be desirable to maximize the surface area of a chemically active membrane within a confining volume, and negatively curved molecular chainmail may be a suitable method to achieve that.

The flat semi-alternating sheets display the most complex behavior. The major and middle squared radii grow with a power close to 1, which is consistent with previous simulations of flat-phase membranes. The roughness exponent takes a value just below one-half, which is smaller than in tethered membrane simulations. We note that the fluctuations of the membrane are anisotropic, taking the form of bends along the long axis of the diamond shape of the membrane, with the short axis remaining stable. 
This distinctive anisotropy can be rationalized by the underlying symmetry of the network, which possesses a 4-ring unit cell with mirror planes along one diagonal, as opposed to the non-alternating network, which has mirror planes along both diagonals of the unit cell.
The correspondingly anisotropic shape fluctuations are analogous to the highly anisotropic elasticity that is seen in knitted fabric, which similarly possess mirror planes along a single direction; the semi-alternating sheets in fact possess a linking motif resembling that of the garter stitch~\cite{Singal2024}.
The lack of fluctuations along one direction likely suppresses the roughness exponent. The finite thickness of the network, due in part to the tendency of each ring to lie with a normal in the plane of the membrane, may suppress roughness further. Notably, the minor eigenvalue of the largest membranes grows faster than the trend set by smaller membranes. This happens when the fluctuations of the long diamond axis become sufficiently large as to cause the membrane to fold over itself, displacing the center of mass from within the loops. 
This bears similarity to the thermal crumpling transition predicted for tethered membranes, in which the distribution of surface normals lose correlation above a certain thermal lengthscale~\cite{kovsmrlj2017statistical}.
It also bears similarity to a backfolding transition observed in nanoconfined DNA that occurs when a molecule exceeds its ``global'' persistence length \cite{dorfman}. A video of a fluctuating semi-alternating network may be found in the ancillary data.


To characterize this folding transition, we note that the diamond-shape networks only fold in the direction of the diamond's long axis, leading to much larger fluctuations in the distance between the two corners separated by the long axis ($D_L$), compared to the two corners separated by the short axis ($D_S$). Typically, the short axis is highly aligned with the medium gyration eigenvector, the eigenvalue of which displays much lower-amplitude fluctuations than the other two, which are anti-correlated. In extreme folding events, the major eigenvector will become aligned with the short axis of the diamond, as can be seen in Fig.~6. These folds are also marked by a displacement of the center of mass of the network away from the geometric center of the middle loop, which greatly increases the minor gyration radius. We can characterize the scale of these fluctuations by measuring the variance in $D_L$ and $D_S$ and plotting their ratio as a function of system size (Fig.~6b). This ratio increases with the number of links, but increases much more strongly beyond 100 links. There is evidence of a local maximum at 49 loops that may be indicative of a phase transition. 

In contrast to tethered membranes with isotropic interactions that undergo a crumpling transition, the semi-alternating chainmail networks are an anisotropic 2D material, and undergo what can be described as a folding transition rather than a crumpling transition. The thermal lengthscale for crumpling is on the order of $\ell_{\rm th}=\kappa\sqrt{kT/Y}$ where $\kappa$ is the bending modulus and $Y$ is the Young's modulus. Although beyond the scope of this work, future simulations applying uniaxial extension or clamped boundaries may measure these moduli to further clarify the lengthscale associated with the folding transition.


\begin{figure}
    \centering
    \includegraphics[width=1\textwidth]{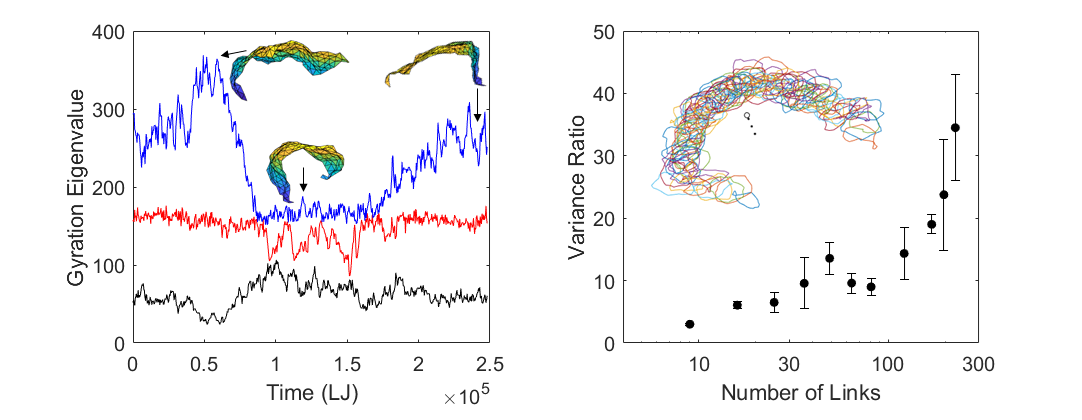}
    \caption{Left: a partial time series of the three gyration eigenvalues of a 225-loop semi-alternating membrane during which a large-amplitude folding event occurs, causing the major axis to align with the width of the diamond-shaped membrane. Delaunay triangulations of the membranes are shown from the side at three stages of this event. Right: ratio of the variance of the distances between the corners along the long axis of the membrane to the variance of the short axis. Error bars represent standard error over multiple runs. Inset shows a side view of a folded membrane, showing its center of mass as a black circle displaced from the loops, and smaller dots pointing along the minor eigenvector.}
    \label{fig:flucto}
\end{figure}


In the two previous kinetoplast simulation studies, the minicircles were rigid or effectively rigid: Polson, Garcia, and Klotz simulated rigid circles~\cite{polson2021flatness}, and He et al.~simulated rings with a contour length half of the persistence length~\cite{he2023single}. As mentioned, the minicircles in \textit{T.~brucei} kDNA are about 1000 base pairs or six persistence lengths, while in \textit{Crithidia fasciculata} they are about 2500 base pairs or 16 persistence lengths. We will briefly comment on the effect of changing the contour and persistence length of the simulated polymer model, as the effects of chirality are the main focus of this manuscript. Alternating networks can pucker when the persistence length is increased, similar to puckering observed by He et al.~\cite{he2023single}. Fully non-alternating networks lose some isotropy and become more taco-shaped. Semi-alternating networks remain diamond-shaped. When the persistence length is kept the same but the loops are increased to 80 beads to mimic \textit{Crithidia} minicircles, we do not see curvature effects as strongly. This is likely due to the reduced exlcuded volume interactions between the rings, and we expect that sufficiently large networks would revert to the established trends.

\subsection{Constrained Gradient Optimization}

We annealed towards the ideal configurations of the three square lattice chiralities with 9, 16, and 25 loops. Larger networks were attempted but did not converge on a minimal configuration, even with more vertices per loop. As a preliminary investigation, we annealed polycatenanes with varying degrees of twist to measure the effect of intrinsic twist on ropelength, described in the Appendix.

\begin{figure}
    \centering
    \includegraphics[width=0.9\textwidth]{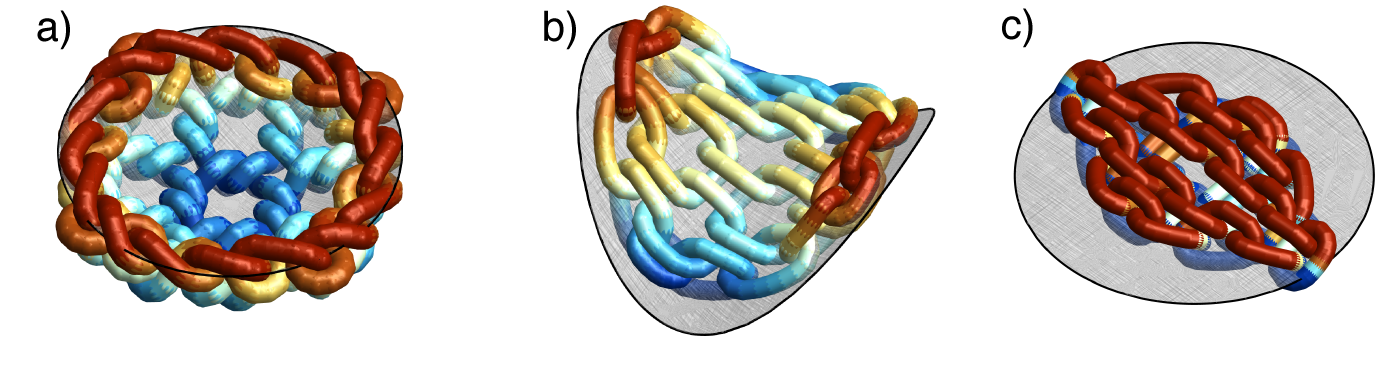}
    \caption{Renderings of the tightest configurations of (from left to right) alternating, fully non-alternating, and semi-alternating chainmail with 25 links, with osculating surfaces. Links are colored by their position along the surface normal direction.}
    \label{fig:tightnets}
\end{figure}

We begin with a qualitative description of the ideal networks. Examining the tight configurations in Fig.~\ref{fig:tightnets}, the alternating, non-alternating, and semi-alternating networks display positive, negative, and null Gaussian curvature as in the polymer simulation. The semi-alternating case in particular remains effectively flat with increasing size. Quantitatively, we can measure the total contour length of the network, in addition to the radii of gyration and the Gaussian curvature.

The lower bound on the ropelength of a linked network follows a result from Cantarella et al.~\cite{cantarella2002minimum} that each loop in the network that is Hopf linked to $Q$ neighbors cannot be shorter than a curve that is parallel (and one unit separated) to the minimum convex hull around $Q$ unit-radius disks. While the minimum convex hull is a challenging optimization problem when $Q$ is large, each link in square lattice molecular chainmail has either 2, 3, or 4 linked neighbors. The ideal divalent link is a stadium curve of length $4\pi + 4$, the ideal trivalent link is a rounded triangle with length $4\pi + 6$, and the ideal tetravalent link is a rounded square or rhombus with a length of $4\pi + 8$. A square network with $M=L^2$ components will have $(L-2)^2$ tetravalent links in the interior, $4(L-2)$ trivalent links on the edges, and 4 divalent links in the corners. For networks of side length $L = 3$, 4, 5, this gives lower bounds of approximately 161.1, 297.1, and 474.2. 

We can compare the total contour length of each network to the lower bound. In all cases, the excess contour increases with the number of links. The alternating network exceeds it by the most, by 31\%, 39\%, and 47\% with 9, 16, and 25 links. The non-alternating exceeds the minimum by 7\%, 13\%, and 29\%, and the semi-alternating is the closest to the ideal configuration, exceeding it by 4\%, 6\%, and 12\%. It may be argued that the larger networks are simply not being annealed as efficiently, but it is likely that the loops on the edge do not need to twist as much to accommodate the topological constraints, and larger networks have a larger proportion of interior loops. The mean and Gaussian curvature of each network can be compared to the expected behavior of surfaces. Here, we expect the ratio of Gaussian curvature to the square of mean curvature to be unitary for surfaces of positive Gaussian curvature, to diverge to negative infinity for surfaces with negative Gaussian curvature, and zero for flat surfaces. Examining the 25-link tightened networks in Fig.~\ref{fig:tightnets}, we find values of 0.999982, -201.824 and 0.779 respectively. We note that the curvature of a tight configuration does not necessarily dictate the equilibrium conformation of a polymer model of the same network. For example, a Japanese-style square network can be constructed in a plane out of ropelength-minimizing components, but maintains positive Gaussian curvature as a polymer (Fig.~\ref{fig:japanborro}).


Although the annealed networks approach the lower bound of ropelength to within tens of percent, it is clear from visual inspection that many of the configurations have not reached their true minimum. This is likely due to the fact that Ridgerunner is optimized for tightening sections of densely linked curves, but not for reducing the length of the straight segments of each link. While these results do highlight the role of topological constraints in the chirality-curvature relationship, they remain qualitative.

\begin{figure}
    \centering
    \includegraphics[width=1\textwidth]{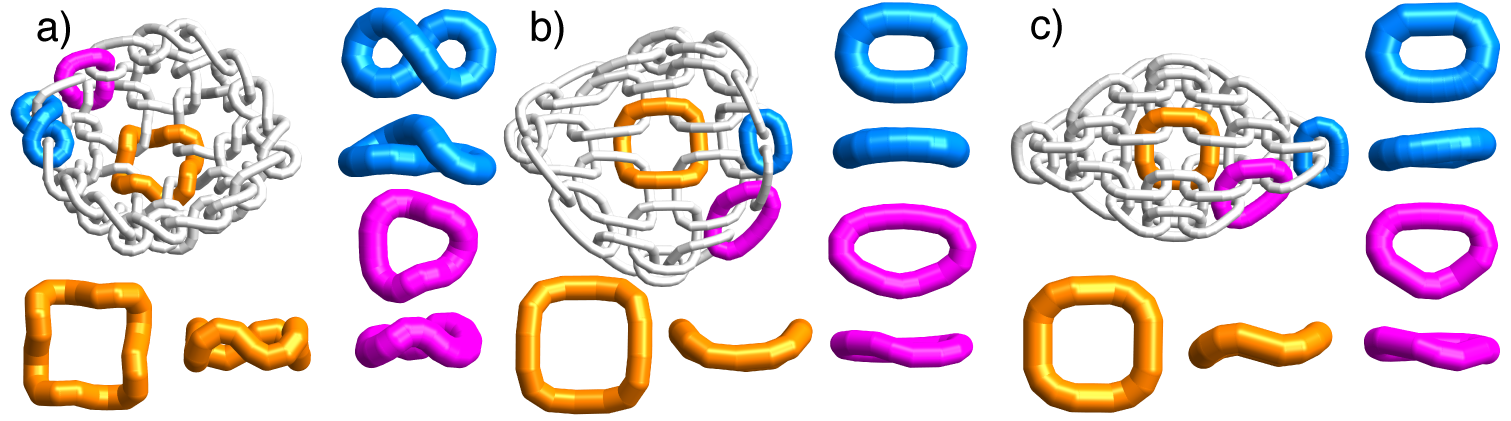}
    \caption{Individual links in tightened networks from the interior (orange), edge (magenta), and corner (blue) of networks with alternating (a), non-alternating (b), and semi-alternating (c) chiralities.}
    \label{fig:tightrings}
\end{figure}

In addition to the overall properties of the tight networks, we can examine the tightest shape of the links from the interior, edges, and corners of the networks. Examples for each position and chirality are seen in Fig.~\ref{fig:tightrings}. The out-of-plane deviation of each ring must accommodate its neighbors according to the chirality, with the flat networks remaining the most planar. The divalent corner loops in the alternating case appear to be twisted into a figure-8 shape, which is also observed in the corner loops in the networks simulated using Langevin Dynamics. Although kinetoplast minicircles are typically not supercoiled, this indicates how an ``open circular'' molecule may have a twisted shape due to external topological constraints.

\section{Link chirality and Gaussian curvature}

\begin{figure}
    \centering
    \includegraphics[width=.9\textwidth]{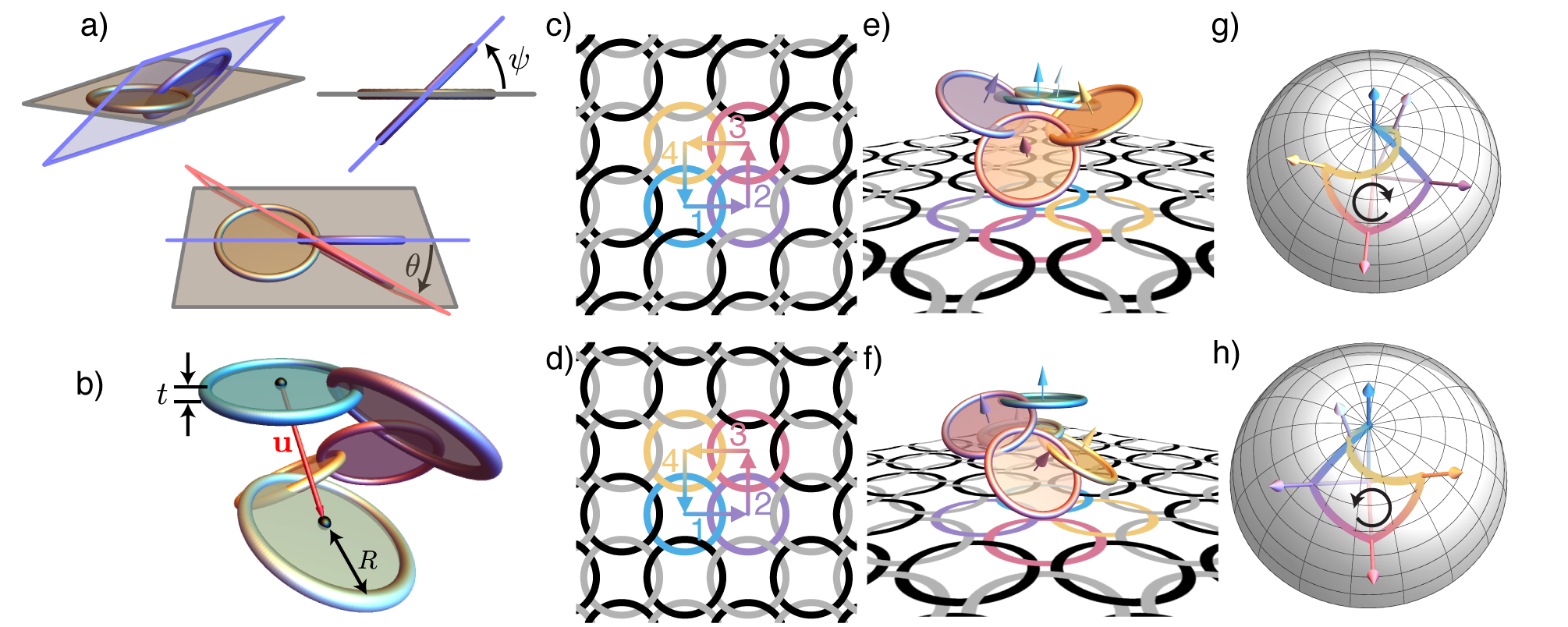}
    \caption{(a) Depiction of the twist angle $\psi$ and bend angle $\theta$ for two linked rings. (b) A non-closed configuration of a chain of linkages, where $R$ is the ring radius, $t$ is the ring thickness, and $\mathbf{u}$ is the vector joining the end of the chain to the beginning. Closed circuits are shown on the diagrams representing the (c) non-alternating chainmail and (d) alternating chainmail, where the order 1-4 shows the counterclockwise ordering of the path. (e) and (f) show these linkage circuits in Euclidean space, where $\psi = \pi/10$ and $\theta$ minimizes $\left|\mathbf{u}\right|^2$, for non-alternating and alternating chainmail, respectively. (g) and (h) show the Gauss maps of these circuits, with clockwise path indicating a negative curvature and counterclockwise path indicating a positive curvature.}
    \label{fig:ring_embedding}
\end{figure}

We now address how linking chirality relates to Gaussian curvature of a chainmail sheet by considering a largely system-agnostic, geometric model of frustrated linkage closure.
In the systems presented so far, the key element is a closed loop of material whose expected configuration is a circular ring of some fixed radius $R$.
In the case of ring polymers, the expected circular conformation is the ensemble-averaged state of a solitary ring polymer; for tight ropes, the circular ring minimizes rope length while maximizing the amount of area that the rope can enclose.
Since each ring has a finite thickness $t$, two rings that are linked together are required to pierce through the interior of the other ring, so that of the $\approx \pi R^2$ of interior area, each linked ring decreases the amount of ``empty area'' by $\sim t^2$, constraining the states available to the rings: ropes cannot be tightened to arbitrarily small lengths and ring polymers experience a reduction in conformational entropy.
Since this threaded area depends on the orientation of the two linked rings, given by the normal unit vector $\hat{\mathbf{N}}$, via a function that increases with $\hat{\mathbf{N}}^1 \cdot \hat{\mathbf{N}}^2$ (here, 1 and 2 index the two linked rings), the optimal configuration is one where the rings are mutually orthogonal.

To describe the geometry of a collection of rings, we first define an orthonormal frame $\{\hat{\mathbf{d}}_1,\hat{\mathbf{d}}_2,\hat{\mathbf{N}}\}$ for a given ring, where $\hat{\mathbf{d}}_\mu$ ($\mu = 1,2$) lie in the plane of the ring.
Arranging these rings on a square lattice, we can then index the rings by a pair of integers $(i,j)$.
If each ring is furthermore assumed to be identical (ignoring boundaries), then we will regard the chainmail sheet as a homogeneous space consisting of ring centers $\mathbf{r}^{(i,j)}$; the normal vector at each point is $\mathbf{N}^{(i,j)}$ and the tangent space at each point is spanned by the vectors $\hat{\mathbf{d}}_\mu^{(i,j)}$.
Next we will orient the tangent space basis vectors such that $\hat{\mathbf{d}}_1^{(i,j)}$ is parallel to $\mathbf{r}^{(i+1,j)} - \mathbf{r}^{(i,j)}$ and $\hat{\mathbf{d}}_2^{(i,j)}$ is parallel to $\mathbf{r}^{(i,j+1)} - \mathbf{r}^{(i,j)}$, both in the limit that $R \to 0$.
Then we define an \emph{discrete affine connection} that relates the frame at two neighboring rings via the rotation matrices $\mathcal{R}_\mu(\theta_\mu,\psi_\mu)$ (provided in the Appendix), where $\theta_\mu$ is a bending angle and $\psi_\mu$ is a twisting angle that each rotate tangent plane vectors into the normal vector according to Fig.~\ref{fig:ring_embedding}(a); rotations about a fixed normal do not play a role.
Next, consider a situation in which $t \ll R$ and linked rings are in contact at their boundaries.
Then neighboring ring centers can be related via
\begin{equation}\label{eq:ring_centers}
\begin{split}
\mathbf{r}^{(i+1,j)} - \mathbf{r}^{(i,j)} &\approx \frac{R}{2}\left(\hat{\mathbf{d}}_1^{(i,j)} + \hat{\mathbf{d}}_1^{(i+1,j)}\right) \\
\mathbf{r}^{(i,j+1)} - \mathbf{r}^{(i,j)} &\approx \frac{R}{2}\left(\hat{\mathbf{d}}_2^{(i,j)} + \hat{\mathbf{d}}_2^{(i,j+1)}\right)
\end{split} \,
\end{equation}
and the configuration can thus be constructed given a set of bending and twisting angles.

\begin{figure}
    \centering
    \includegraphics[width=.7\textwidth]{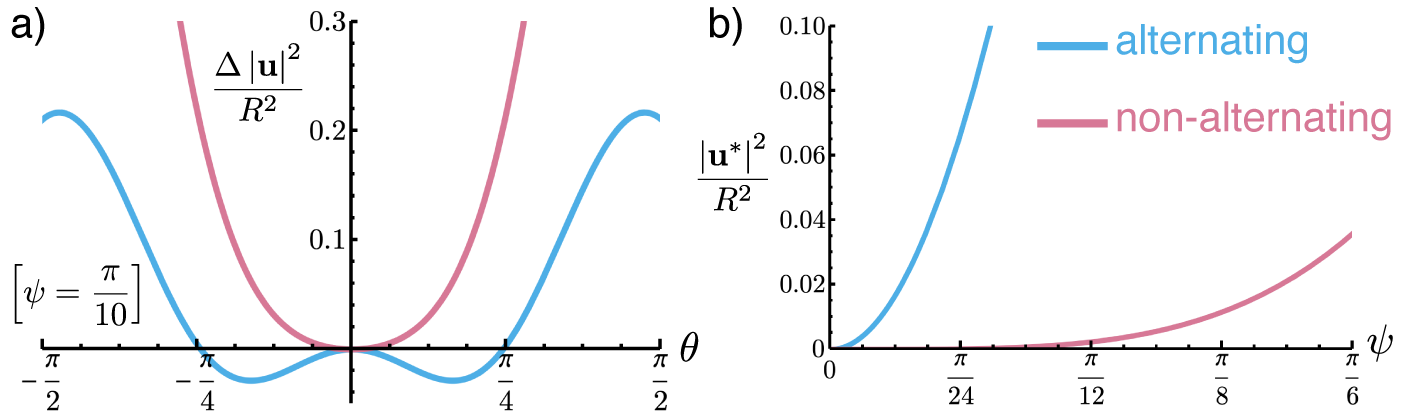}
    \caption{(a) compares the degree of nonclosure, expressed as $\Delta|\mathbf{u}|^2 \equiv (\left|\mathbf{u}(\theta,\psi)\right|^2-\left|\mathbf{u}(0,\psi)\right|^2)/R^2$, as a function of bend angle $\theta$ for twist angle $\psi = \pi/10$. (b) compares the minimum value of the non-closure $\left|\mathbf{u}^*\right|^2/R^2 \equiv \left|\mathbf{u}(\theta^*(\psi),\psi)\right|^2/R^2$ as a function of twist angle $\psi$.}
    \label{fig:displacement_plots}
\end{figure}

Given the affine connection and \ref{eq:ring_centers}, we can then assess the degree to which the chainmail sheet can be approximated by a smooth, homogeneous manifold embedded in Euclidean space.
We do this by examining a closed circuit along the lattice -- a path $\mathbf{r}^{(i,j)}\mapsto \mathbf{r}^{(i+1,j)} \mapsto \mathbf{r}^{(i+1,j+1)} \mapsto \mathbf{r}^{(i,j+1)} \mapsto \tilde{\mathbf{r}}^{(i,j)}$ -- and determining the magnitude of the displacement vector $\mathbf{u}$ required to join the ring center $\mathbf{r}^{(i,j)}$ (the origin of the circuit) to the ring center $\tilde{\mathbf{r}}^{(i,j)}$ (the end-point of the circuit), as depicted in Fig.~\ref{fig:ring_embedding}(b).
The lattice circuits for non-alternating and alternating chainmail are shown in Figs.~\ref{fig:ring_embedding}(c) and (d).
Lattice symmetry dictates the values of the bending angles $\theta_\mu$ and twisting angles $\psi_\mu$, such that for the non-alternating lattice, $\theta_2 = -\theta_1 \equiv -\theta$ and $\psi_2 = -\psi_1 \equiv -\psi$, and for the alternating lattice, $\theta_2 = \theta_1 \equiv \theta$ and $\psi_2 = \psi_1 \equiv \psi$.
Since vanishing values of the twist angle $\psi$ result in rings whose boundaries pass through each other, thus requiring rings to deviate from their preferred circular shape, we fix $\psi$ to be non-zero and then consider values of $\theta$ that minimize $\left|\mathbf{u}\right|^2$.
The case of $\psi = \pi/10$ is shown in Figs.~\ref{fig:ring_embedding}(e) and (f) for non-alternating ($\theta = 0$) and alternating $(\theta \approx 0.513)$ chainmail, respectively, revealing that these linkages do not generally close when immersed in Euclidean space.
At fixed $\psi$, the magnitude of non-closure is shown as a function of bending angle $\theta$ in Fig.~\ref{fig:displacement_plots}(a).
While non-alternating sheets minimize their non-closure by twisting without bend ($\theta = 0$), alternating sheets adopt one of two non-zero bend angles $\pm\theta^*(\psi)$.
In a small-angle approximation, we find for alternating sheets that the degree of non-closure is given by
\begin{equation}
    \frac{\Delta\left|\mathbf{u}(\theta,\psi)\right|^2}{R^2} \equiv \frac{\left|\mathbf{u}(\theta,\psi)\right|^2 - \left|\mathbf{u}(0,\psi)\right|^2}{R^2} \approx -\frac{5}{2}\psi^2\theta^2 + \frac{1}{2}\theta^4 \, ,
\end{equation}
so that the optimal bending angle is  $\theta^*(\psi) \approx \pm \sqrt{\frac{5}{2}}\psi$.

Appealing to the Volterra construction of defects in elastic media, each loop in the lattice may be regarded as enclosing a dislocation core and the closure of these loops requires an elastic deformation from the preferred state, the cost of which scales with $\left|\mathbf{u}\right|^2$~\cite{ChaikinLubensky,Kleman2008}.
Since the amount of dislocation grows with the number of rings enclosed by the lattice circuit, the elastic cost grows super-extensively, requiring increasingly large deformations at the boundary of the sheets, thus limiting their maximum sizes~\cite{Hagan2021,Efrati2021}, and resulting in large differences in the shapes of bulk and boundary loops, as illustrated in Fig~\ref{fig:tightrings}.
As shown in Fig.~\ref{fig:displacement_plots}(b), the cost of elastic deformation is smaller for non-alternating sheets, increasing as $\left|\mathbf{u}^*(\psi)\right|^2/R^2 \approx \frac{1}{2}\psi^4$, whereas for alternating sheets, $\left|\mathbf{u}^*(\psi)\right|^2/R^2 \approx 4\psi^2 - \frac{95}{24}\psi^4$.
Consequently, alternating sheets are under higher residual stress when forced to maintain closed links in Euclidean space.

Finally, we address the expected curvatures of these sheets by studying the transport of the ring frame around a lattice circuit.
The curvature is related to the net rotation of the frame around a closed circuit, which is given by the commutator of $\mathcal{R}_1$ and $\mathcal{R}_2$, namely $\mathcal{R}_\square \equiv \mathcal{R}_2^{-1}\mathcal{R}_1^{-1}\mathcal{R}_2\mathcal{R}_1$.
Taking the small-angle approximation, we find that
\begin{equation}\label{eq:circuit_commutator_result}
    \mathcal{R}_\square \approx \left(\begin{array}{ccc}
    1 & -(\theta_1\theta_2 + \psi_1\psi_2) & 0 \\
    (\theta_1\theta_2 + \psi_1\psi_2) & 1 & 0 \\
    0 & 0 & 1
    \end{array}\right) \, ,
\end{equation}
so to leading order in the bending and twisting angles, the frame undergoes an in-plane rotation when transported around a loop, indicating that the Gaussian curvature $K$ is given by
\begin{equation}
    K_{\rm non-alt.} \approx -\frac{\psi^2}{R^2} \, 
\end{equation}
for non-alternating chainmail and
\begin{equation}
    K_{\rm alt.} \approx \frac{7\psi^2}{2R^2}
\end{equation}
for alternating chainmail.
Therefore, we confirm that non-alternating chainmail is characterized by negative Gaussian curvature (i.e.~is hyperbolic) and alternating chainmail is characterized by positive Gaussian curvature (i.e.~is spherical).
The identification of the alternating chainmail as spherical allows us to interpret the two preferred bending angles as corresponding to the two choices sign for the mean curvature $H \approx \pm \sqrt{5/2}\psi/R$.
Beyond the small-angle approximation, we additionally consider transport of the normal vector $\hat{\mathbf{N}}$ around the lattice circuit.
The unit vectors are mapped to the unit two-sphere $\mathbbm{S}^2$ via the Gauss map, and the individual ring normals form the corners of the paths shown in Fig.~\ref{fig:ring_embedding}(g) and (h) for non-alternating and alternating chainmail, where the segments of the path are found using linear interpolation.
The solid angle formed by the paths on $\mathbbm{S}^2$ is proportional to the integrated Gaussian curvature of the surface patch that they represent, and the orientation of the path is controlled by the sign of the Gaussian curvature, providing a secondary confirmation that the two chainmail sheets have different curvatures~\cite{Needham2021}.
Notably, these paths do not close, indicating a net rotation of the normal that is not found in the small angle approximation.
This suggests a net twist or torsion of the frame as it is transported around the circuit.
Since surfaces are not allowed to have torsion, this is another source of stress that must be accounted for by ring deformation in order for the sheet to be embedded in Euclidean space.

\section{Conclusion}

The most significant result of this work is a set of simulation data showing the connection between the chirality of links in molecular chainmail and the Gaussian curvature of the surface. We have demonstrated this both for a DNA-parameterized polymer model, and for geometrically tight ideal configurations. We have demonstrated that networks that are flat at small molecular weights will undergo a transition to folded configurations above a certain size, analogous to the crumpling transitions predicted for tethered membranes.

Analogous systems are subject to theorems relating curvature to chirality in some way. The Fuller-White-Calugareanu theorem relates the topology and geometry of a twisted ribbon (or a DNA double-helix), constraining the sum of twist and writhe of the ribbon to the linking number of its two edges (or helices). Likewise, the Gauss-Bonnet theorem relates the Gaussian curvature of a surface to its donut-hole genus. An analogous theorem may be developed to relate the chirality of a linked surface to its Gaussian curvature. Drawing on the similarity between the polymer and tight networks in Figs.~\ref{fig:postlammps} and \ref{fig:tightnets}, a constraint may be established (for example) between the Mobius energy of the links and the Helfrich energy of the surface. 
As a first step towards such a relation, we have developed an understanding of the topology-curvature relation through a model system of linked rigid rings.
This purely kinematic construction, which shares similarities with geometric relations for periodic origami~\cite{McInerney2020}, defines a discrete affine connection based on the constraints introduced by the configuration of Hopf links between pairs of rings.
Intrinsic curvature is then derived from the holonomy of this discrete connection.
Since this construction is purely based on local rules, without \emph{a priori} postulating a specific manifold structure, we find predictions for rings that preferentially lie in defect-ridden lattices, suggesting that a geometry-curvature relationship for periodically-linked rings in the spirit of Fuller-White-Calugareanu may resemble the relationships between topological defects and curvature of flexible crystalline membranes~\cite{Seung1988}.
Moreover, the observed relationships between link symmetry and membrane shape draws parallels with similar relationships between stitch topology and fabric shape and mechanics in the context of knitted fabric~\cite{Singal2024}, which we leave for future explorations.
We hope that this work allows the discovery of a more complete understanding of the topology-curvature relation for surfaces, and guides the rational design of future planar materials with topologically complex chemistry.

\section{Acknowledgements}
ARK's work is supported by the National Science Foundation, grant number 2122199. We are grateful to Cristian Micheletti, Juan Luengo-M\'arquez, and Salvatore Assenza for sending us an advanced copy of their manuscript.


\bibliographystyle{unsrt}
\bibliography{linkrefs}

\section*{Appendix: Tight Twisted Polycatenanes and Bonus Networks}

Here we discuss the results of constrained gradient optimization of twisted polycatenanes using Ridgerunner. For an introduction to twisted polycatenanes we direct the reader to Tubiana et al.~\cite{tubiana2022circular}. We initialized polycatenanes as circular networks of 22-gons, each with two possible link chiralities. The polycatenanes had an even number of links. If all links have the same chirality, the network is fully non-alternating and untwisted. If a link had the opposite chirality of its two neighbors, it gives the network alternating crossings and excess twist. A network is said to be maximally twisted if half the links contribute to twisting the network. When the polycatenane is fully non-alternating and untwisted, each link may be minimized to a stadium curve with ropelength $4\pi+4$~\cite{cantarella2002minimum}, and the network takes on a familiar chain-link configuration with a ropelength of $M(4\pi+4)$. The simplest links that admit this pattern are the 4-component $8^4_3$, $8^4_2$, and $8^4_1$ with zero, one, and two twists.

\begin{figure}
    \centering
    \includegraphics[width=\textwidth]{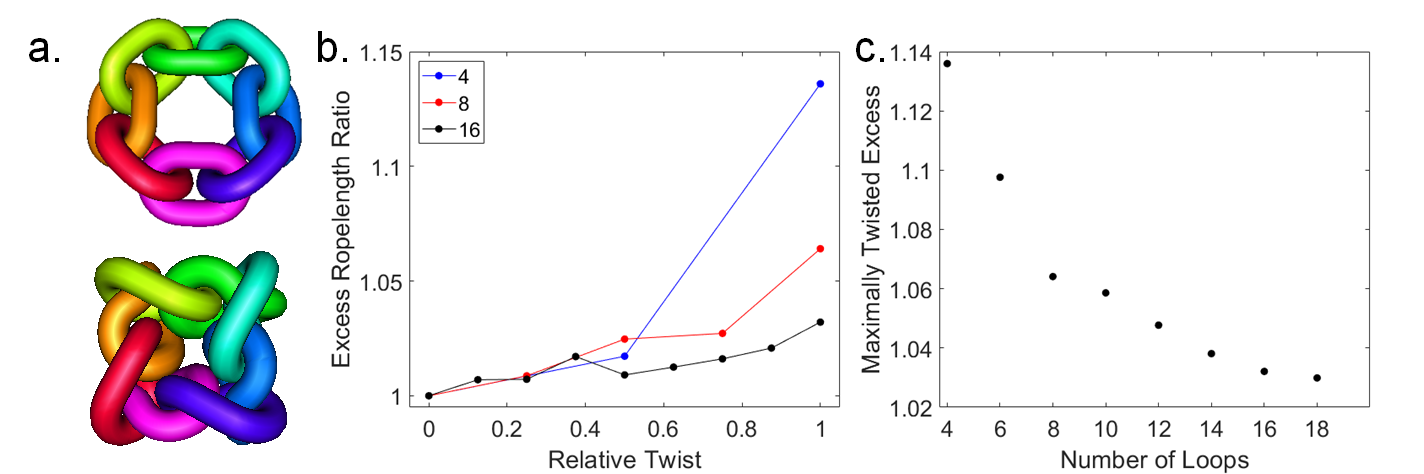}
    \caption{a. Ideal configurations of untwisted and maximally twisted 8-loop polycatenanes. b. Excess ropelength ratio of ideal twisted polycatenanes, as a function of their relative twist as defined in the text. c. Excess ropelength ratio of maximally twisted polycatenanes as a function of the number of loops.}
    \label{fig:appendix}
\end{figure}

We measured the ropelength of twisted polycatenanes with $M$ from 6 to 18 loops, varying the twist from zero to $M/2$. Untwisted polycatenanes formed circular chain-link configurations as expected, and as the twist was increased, each component in the network became twisted. The final configuration often picked up the symmetry of a regular $M/2$-gon, for example the 8-loop polycatenane forming a square (Fig.~\ref{fig:appendix}a). The process often involved a buckling in which the normal vector of each loop was forced to rotate $180^\circ$. Since the ropelength depends more strongly on the number of loops than their twist, we plot the ratio of the measured ropelength to the untwisted minimum, against the degree of twist $\frac{N_{tw}}{M/2}$ ranging from 0 to 1. Fig.~\ref{fig:appendix}b shows this data for polycatenanes with 4, 8, and 16 loops. Although the excess ropelength similarly increases with twist for each sized network, the maximum excess ropelength at maximum twist is largest for the smallest networks. The excess ropelength at maximal twist as a function of the number of loops in the network is plot in Fig.~\ref{fig:appendix}c, where it decreases roughly as $1+2/M$. The tightest knot is not necessarily the smallest~\cite{klotz2022tightest}, and the maximally twisted polycatenanes, despite having a greater ropelength, occupied a smaller volume as defined by their convex hull. This may have implications for the evolution of supercoiling in bacterial DNA.

In a square lattice chainmail network, each set of $3 \times 3$ loops maybe treated as an 8-loop polycatenane constrained by a ninth in the middle. We expect alternating networks to have loops that are more deformed from their ideal configuration than the comparable maximally-twisted 8-loop polycatnenaes in Fig.~\ref{fig:appendix}a, acquiring at least 6\% excess ropelength. Although direct comparisons between polycatenanes and chainmail networks are difficult, this was a necessary first step towards our current understanding.

\begin{figure}
    \centering
    \includegraphics[width=\textwidth]{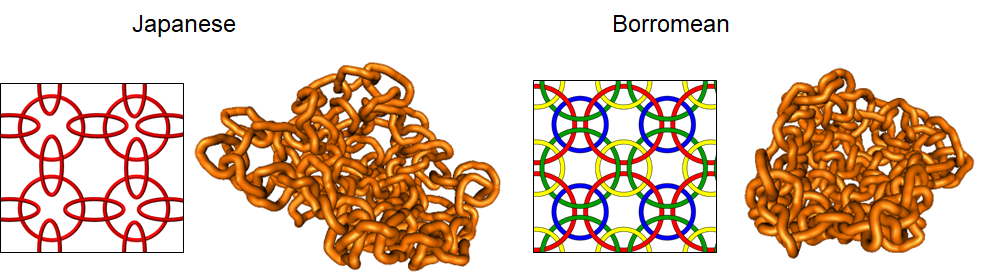}
    \caption{Design and LAMMPS configurations of Japanese and Borromean chainmail.}
    \label{fig:japanborro}
\end{figure}

Figure \ref{fig:japanborro} shows the patterns and sample LAMMPS configurations of Japanese 4-in-1 chainmail, studied by Polson et al.~\cite{polson2021flatness}, and Borromean chainmail, based on a design by Luc Devroye~\cite{devroye}. Japanese chainmail has no chirality effects, and no two links in Borromean chainmail share a direct topological link. We have not done quantitative analysis on either, but both adopt positive Gaussian curvature. The Japanese chainmail network can be constructed out of ropelength-minimizing components in a plane, implying that an ideal configuration need not have the same curvature as a polymer configuration.

\section*{Appendix: Quadric surface fitting}

To perform the quadric surface fitting, we first calculated the center of mass $\mathbf{r}^*_{i}$ for each ring $i$ and the center of mass $\langle\mathbf{r}\rangle$ for the full chainmail assembly.
We then calculated the gyration tensor $G_{ab}$, defined in Eq.~\ref{eq:gyration} and found the normalized eigenvectors $\left\{\hat{\mathbf{e}}_1,\hat{\mathbf{e}}_2,\hat{\mathbf{e}}_3\right\}$, ordered from largest to smallest eigenvalue.
The quadric surface is then given by the set of points $\bm{\sigma}$, expressed using the Monge representation as
\begin{equation}
\bm{\sigma}(u_1,u_2) = \langle \mathbf{r} \rangle + u_1\hat{\mathbf{e}}_1 + u_2\hat{\mathbf{e}}_2 + \zeta(u_1, u_2)\hat{\mathbf{e}}_3\, ,
\end{equation}
where the Monge height function $\zeta(u,v)$ is given by
\begin{equation}
\zeta(u_1,u_2) = \frac{1}{2}u_{\mu}C_{\mu \nu}u_{\nu} + B_\mu u_\mu + A_0\, ,
\end{equation}
where $A_0$ is a scalar, $B_\mu$ are components of a vector $\mathbf{B}$, and $C_{\mu\nu}$ are components of a symmetric matrix $\mathbf{C}$ (where $C_{12} = C_{21}$).
The quadric surface is then fit to ring center of mass data by minimizing the residual functional 
\begin{equation}
I[A_0, \mathbf{B}, \mathbf{C}] = \frac{1}{M}\sum_{i=1}^M\left(r^*_{i,3} - \zeta(r^*_{i,1},r^*_{j,2})\right)^2
\end{equation}
over the parameters $A_0$, $\mathbf{B}$, and $\mathbf{C}$.
Here, the ring centers $\mathbf{r}^*_i$ have been expressed in the $\left\{\hat{\mathbf{e}}_1,\hat{\mathbf{e}}_2,\hat{\mathbf{e}}_3\right\}$ frame as $\mathbf{r}^*_i = r^*_{i,a}\hat{\mathbf{e}}_a$.
Finally, the Gaussian curvature $K$ is approximated as
\begin{equation}
    K \approx \frac{{\rm det}\, \mathbf{C}}{\left(1 + |\mathbf{B}|^2\right)^2}
\end{equation}
and the mean curvature $H$ is approximated as
\begin{equation}
    H \approx \frac{(1 + B_2^2)C_{11} - 2 B_1 B_2 C_{12} + (1 + B_1^2)C_{22}}{2\left(1 + |\mathbf{B}|^2\right)^{3/2}}\, .
\end{equation}

\section*{Appendix: Discrete affine connection}

The discrete affine connection is a pair of rotation matrices $\mathcal{R}_1$ and $\mathcal{R}_2$.
The first of these, $\mathcal{R}_1$, maps the orthonormal frame at ring $(i,j)$ to the frame at $(i+1,j)$ via
\begin{equation}\label{eq:affine_connection_1}
\left(\begin{array}{c}
\hat{\mathbf{d}}_1^{(i+1,j)} \\
\hat{\mathbf{d}}_2^{(i+1,j)} \\
\hat{\mathbf{N}}^{(i+1,j)} \\
\end{array}\right) = 
\left(\begin{array}{ccc}
\cos\theta_1 & -\sin\theta_1\sin\psi_1 & -\sin\theta_1\cos\psi_1 \\
0 & \cos\psi_1 & -\sin\psi_1 \\
\sin\theta_1 & \cos\theta_1\sin\psi_1 & \cos\theta_1\cos\psi_1
\end{array}\right)\left(\begin{array}{c}
\hat{\mathbf{d}}_1^{(i,j)}\\
\hat{\mathbf{d}}_2^{(i,j)} \\
\hat{\mathbf{N}}^{(i,j)} \\
\end{array}\right)
= \mathcal{R}_1(\theta_1,\psi_1)\left(\begin{array}{c}
\hat{\mathbf{d}}_1^{(i,j)}\\
\hat{\mathbf{d}}_2^{(i,j)} \\
\hat{\mathbf{N}}^{(i,j)} \\
\end{array}\right)
\end{equation}
The second, $\mathcal{R}_2$, maps the orthonormal frame at ring $(i,j)$ to the frame at $(i,j+1)$ via
\begin{equation}\label{eq:affine_connection_2}
\left(\begin{array}{c}
\hat{\mathbf{d}}_1^{(i,j+1)} \\
\hat{\mathbf{d}}_2^{(i,j+1)} \\
\hat{\mathbf{N}}^{(i,j+1)} \\
\end{array}\right) = 
\left(\begin{array}{ccc}
\cos\psi_2 & 0 & \sin\psi_2 \\
\sin\theta_2\sin\psi_2 & \cos\theta_2 & -\sin\theta_2\cos\psi_2 \\
-\cos\theta_2\sin\psi_2 & \sin\theta_2 & \cos\theta_2\cos\psi_2
\end{array}\right)\left(\begin{array}{c}
\hat{\mathbf{d}}_1^{(i,j)}\\
\hat{\mathbf{d}}_2^{(i,j)} \\
\hat{\mathbf{N}}^{(i,j)} \\
\end{array}\right)
=\mathcal{R}_2(\theta_2,\psi_2)\left(\begin{array}{c}
\hat{\mathbf{d}}_1^{(i,j)}\\
\hat{\mathbf{d}}_2^{(i,j)} \\
\hat{\mathbf{N}}^{(i,j)} \\
\end{array}\right)
\end{equation}
Taking the small-angle approximation, we $\mathcal{R}_\mu \approx \mathbbm{1} + \mathcal{A}_\mu$, where
\begin{equation}\begin{split}
\mathcal{A}_1 &= \left(\begin{array}{ccc}
0 & 0 & -\theta_1 \\
0 & 0 & -\psi_1 \\
\theta_1 & \psi_1 & 0
\end{array}\right) \\
\mathcal{A}_2 &= \left(\begin{array}{ccc}
0 & 0 & \psi_2 \\
0 & 0 & -\theta_2 \\
-\psi_2 & \theta_2 & 0
\end{array}\right)
\end{split}\end{equation}
which allows us to approximate the resultant rotation matrix $\mathcal{R}_\square$ around the closed circuit as
\begin{equation}
    \mathcal{R}_\square \approx \mathbbm{1} + \left[\mathcal{A}_2, \mathcal{A}_1\right]\, ,
\end{equation}
yielding the result in Eq.~\ref{eq:circuit_commutator_result}.
Here, the commutator $\left[\mathcal{A}_2, \mathcal{A}_1\right] \equiv \mathcal{A}_2\mathcal{A}_1 - \mathcal{A}_1 \mathcal{A}_2$ measures the holonomy of the discrete connection and represents a discrete calculation of the curvature two-form~\cite{Needham2021}.
Note that the only non-zero component of the commutator is a generator of rotations that leave the normal fixed, allowing us to immediately relate the magnitude of this term to the Gaussian curvature.

\end{document}